\documentclass[conference]{IEEEtran}
\IEEEoverridecommandlockouts
\usepackage{cite}
\usepackage{amsmath,amssymb,amsfonts}
\usepackage{algorithmic}
\usepackage{graphicx}
\usepackage{textcomp}
\usepackage{xcolor}
\usepackage{makecell} 
\usepackage{url} 
\def\BibTeX{{\rm B\kern-.05em{\sc i\kern-.025em b}\kern-.08em
    T\kern-.1667em\lower.7ex\hbox{E}\kern-.125emX}}
\begin{document}

\title{Advances in Battery Energy Storage Management: Control and Economic Synergies\\

\thanks{Funded by Southwest Research Institute}
}

\author{\IEEEauthorblockN{1\textsuperscript{st} Venkata Rajesh Chundru}
\IEEEauthorblockA{\textit{Powertrain Engineering} \\
\textit{Southwest Research institute}\\
San Antonio, Texas \\
venkata.chundru@swri.org}
\and
\IEEEauthorblockN{2\textsuperscript{nd} Shreshta Rajakumar Deshpande}
\IEEEauthorblockA{\textit{Powertrain Engineering} \\
\textit{Southwest Research institute}\\
San Antonio, Texas \\
shreshta.rajakumardeshpande@swri.org} 
\and
\IEEEauthorblockN{3\textsuperscript{rd}  Stanislav A Gankov}
\IEEEauthorblockA{\textit{Powertrain Engineering} \\
\textit{Southwest Research institute}\\
San Antonio, Texas \\
stas.gankov@swri.org}
}

\maketitle
\pagestyle{plain}
\pagenumbering{arabic}

\begin{abstract}
The existing literature on Battery Energy Storage Systems (BESS) predominantly focuses on two main areas: control system design aimed at achieving grid stability and the techno-economic analysis of BESS dispatch on power grid. However, with the increasing incorporation of ancillary services into power grids, a more comprehensive approach to energy management systems is required. Such an approach should not only optimize revenue generation from BESS but also ensure the safe, efficient, and reliable operation of lithium-ion batteries. This research seeks to bridge this gap by exploring literature that addresses both the economic and operational dimensions of BESS. Specifically, it examines how economic aspects of grid duty cycles can align with control schemes deployed in BESS systems. This alignment, or synergy, could be instrumental in creating robust digital twins—virtual representations of BESS systems—that enhance both grid stability and revenue potential.

The literature review is organized into five key categories: (1) ancillary services for BESS, exploring support functions that BESS can provide to power grids; (2) control systems developed for real-time BESS power flow management, ensuring smooth operations under dynamic grid conditions; (3) optimization algorithms for BESS dispatch, focusing on efficient energy allocation strategies; (4) techno-economic analyses of BESS and battery systems to assess their financial viability; and (5) digital twin technologies for real-world BESS deployments, enabling advanced predictive maintenance and performance optimization. This review will identify potential synergies, research gaps, and emerging trends, paving the way for future innovations in BESS management and deployment strategies.
\end{abstract}


\section{Introduction}
The global energy landscape is undergoing a profound transformation, driven by the urgent need to decarbonize and the corresponding proliferation of variable renewable energy sources (RES) such as solar photovoltaics (PV) and wind power \cite{b1}. This transition, however, introduces significant challenges to the stability, reliability, and security of traditional power grids, which were designed around predictable, dispatchable fossil fuel-based generation. The inherent intermittency of RES creates fluctuations in power output that can lead to grid imbalances, frequency deviations, and voltage instability \cite{b2} \cite{b3}. In this context, Battery Energy Storage Systems (BESS) have emerged from a niche technology to a cornerstone of modern power systems, representing a critical enabling technology for a high-RES future \cite{b4} \cite{b5}.
BESS offer a versatile and fast-acting solution to the challenges posed by renewables. By absorbing excess energy during periods of high generation and low demand, and injecting it back into the grid during periods of low generation and high demand, BESS can effectively smooth RES output, mitigate curtailment, and enhance grid stability \cite{b6}. The market has recognized this pivotal role, with global BESS deployments projected to experience a tenfold increase between 2022 and 2030, exceeding 400 GWh per year \cite{b5}. In Europe alone, the residential BESS market surpassed the 1 million unit threshold in 2022, indicating robust growth in both behind-the-meter (BtM) and front-the-meter (FtM) applications \cite{b5}. This rapid expansion underscores the industry's consensus that BESS are indispensable for transforming buildings into predictable power sources and ensuring the overall security of the power system.

Despite their growing importance, the academic and industrial research surrounding BESS has historically been characterized by a significant dichotomy, proceeding along two parallel but seldom-intersecting tracks\cite{b7}\cite{b8}. This bifurcation has created a conceptual gap that limits the realization of the full potential of these assets.

The first track has been the operational and control-centric approach. This line of research focuses on the physical asset itself. It encompasses the design of sophisticated control systems, the development of Battery Management Systems (BMS), and the refinement of algorithms for state estimation \cite{b1}. The primary objective of this research stream are to ensure the safe, reliable, and efficient operation of the battery, with a strong emphasis on maximizing its operational lifespan by mitigating degradation mechanisms\cite{b9}. This research is fundamental to the physical integrity and performance of the BESS.

The second track has been the techno-economic and market-centric approach. This research focuses on the BESS as an economic asset participating in electricity markets. It involves techno-economic analyses, studies on revenue generation from various grid services, and the development of dispatch strategies to maximize profitability \cite{b10}. However, to manage the complexity of market modeling, this stream of research often relies on simplified or idealized models of the battery. Degradation, if considered at all, is frequently represented by simple cycle counting, failing to capture the complex, non-linear, and path-dependent nature of battery aging\cite{b10}.

\begin{figure*}[!h]
	\centering
	\includegraphics[scale=0.6]{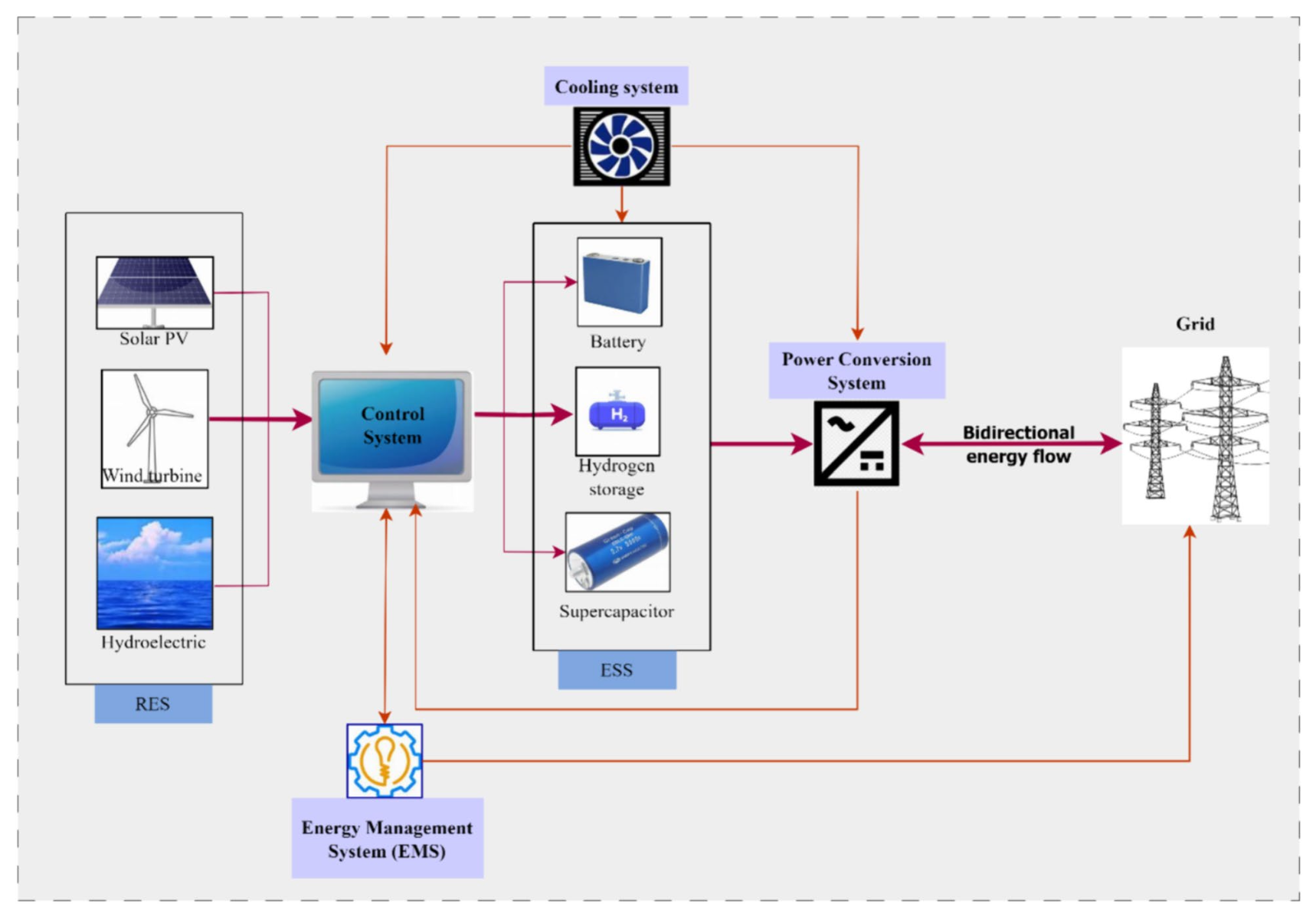}
	\caption{An overview of hybrid energy storage systems and their components \cite{b7}}
	\label{fig:BESS Hybrid Arch}
\end{figure*}

Figure \ref{fig:BESS Hybrid Arch} presents a hybrid architecture of a BESS system where energy management system is balancing the power generation from renewable energy sources and grid side demand actively by controlling BESS charge/discharge cycle. This architecture primarily focuses on BESS safety and health with revenue generation as secondary objective.

This siloed approach, where control engineers focus on battery health and economists focus on market revenue, has become increasingly untenable. Optimizing for market revenue without a high-fidelity understanding of the resulting battery degradation can lead to premature asset failure and destroy economic value. Conversely, operating a battery under overly conservative controls to maximize its life can leave significant revenue on the table, rendering the project financially unviable.

This analytical review focuses on the synergy between techno economic analysis and control system design that can realize the full potential, operational effectiveness, and economic viability of Battery Energy Storage Systems. This synergistic approach would involve holistic integration of high-fidelity operational models and battery degradation—directly into economic optimization and energy management frameworks through energy management system of battery energy storage systems. 

The historical separation of control and economic analysis is a barrier to progress that must be dismantled. The future of advanced BESS management lies in creating a unified system where the physical reality of the battery's health dynamically informs its economic dispatch strategy on a real-time basis. This review will systematically explore the literature across five key domains—ancillary services, control systems, techno-economic analysis, optimization algorithms, and digital twins—to build a comprehensive case for this synergistic paradigm. It will demonstrate how the technical requirements of grid services create specific degradation profiles, how economic incentives are driving the need for more sophisticated algorithms, and how the value gap in BESS profitability can only be closed by co-optimizing for revenue and longevity. Ultimately, this review will culminate in the analysis of Digital Twin (DT) technology as the definitive enabling platform for achieving this essential control-economic synergy, paving the way for the next generation of intelligent, autonomous, and value-optimized energy storage solutions.

\section{Part I: The Operational Mandate: Grid Services and Control Architectures}
This first part of the review establishes the foundational operational context for BESS. Before any economic optimization can be considered, it is essential to understand the technical functions a BESS is expected to perform to support the power grid and the control architectures that ensure its physical integrity. This section addresses the fundamental question of what functions a Battery Energy Storage System (BESS) must perform from both a technical and grid-support perspective.

\subsection{Section 1: The Spectrum of BESS-Provided Ancillary Services}

Battery Energy Storage Systems are uniquely capable of providing a wide array of ancillary services to the grid, thanks to their fast response time, modularity, and ability to both absorb and inject power. The literature consistently categorizes these services based on their technical characteristics, particularly their response time and the duration of energy delivery\cite{b11}\cite{b12}. Understanding this spectrum is the first step in appreciating the complex operational demands placed on BESS assets.

\subsubsection{Categorization of Ancillary Services}
The services provided by BESS can be broadly classified into three main groups:
\begin{itemize}
    \item Short-term fast-response services
    \item Long-term bulk-energy services
    \item overarching grid support and resiliency services.
\end{itemize}

\textbf{Short-Term / Fast-Response Services (Power Applications:} These services are typically deployed over timescales of seconds to minutes and are primarily concerned with maintaining grid stability by managing real-time power balance. Their value lies in the speed and precision of their response.
\begin{enumerate}
\item \textbf{Frequency Regulation and Control:} This is arguably the most prominent and valuable ancillary service provided by BESS \cite{b13}. Power grids must maintain a stable frequency (e.g., 50 Hz or 60 Hz), which requires a constant balance between power generation and load. Deviations from the nominal frequency indicate an imbalance that, if uncorrected, can lead to widespread outages. Due to their ability to respond in milliseconds, BESS are exceptionally effective at injecting or absorbing power to correct these deviations \cite{b14}. This capability allows BESS to participate in specialized, often lucrative, markets such as the Firm Frequency Response (FFR) market in the United Kingdom and the Frequency Control Ancillary Services (FCAS) market in Australia\cite{b14}.

\item \textbf{Voltage Support:} With the increasing penetration of distributed Renewable energy sources (RES) like rooftop solar, local voltage fluctuations in distribution networks have become a significant concern. BESS can provide dynamic voltage support by injecting or absorbing reactive power, helping to maintain voltage profiles within acceptable limits and improving overall power quality \cite{b13}.

\item \textbf{Synthetic Inertia:} Traditionally, the rotational inertia of large synchronous generators in conventional power plants has provided a natural buffer against rapid frequency changes. As these plants are decommissioned and replaced by inverter-based resources like solar and wind, the grid's overall inertia decreases, making it more vulnerable to frequency instability. BESS, through advanced inverter controls, can be programmed to emulate this inertial response, a service known as synthetic or virtual inertia, which is becoming increasingly critical for low-inertia grids \cite{b16}.
\end{enumerate}

\textbf{Long-Term / Bulk-Energy Services (Energy Applications):} These services involve the management and dispatch of energy over longer durations, typically from minutes to several hours. They are focused on optimizing the use of energy over time rather than providing instantaneous power.
\begin{enumerate}
 \item \textbf{Peak Shaving and Load Shifting:} This involves charging the BESS during off-peak hours when electricity demand and prices are low, and discharging it during peak hours when demand and prices are high \cite{b15}. This service provides multiple benefits: it reduces the strain on the grid during peak periods, can defer or eliminate the need for costly network infrastructure upgrades, and can significantly lower electricity bills for end-users by reducing demand charges\cite{b17}\cite{b18}.
 
\item \textbf{Congestion Relief:} In certain parts of a power grid, the physical capacity of transmission or distribution lines can become a bottleneck, limiting the amount of power that can be delivered. This is known as congestion. A strategically placed BESS can alleviate congestion by absorbing power upstream of the constraint and injecting it downstream, effectively bypassing the bottleneck without requiring physical upgrades to the lines\cite{b12}.

\item \textbf{Renewable Energy Smoothing and Curtailment Mitigation:} A major challenge with wind and solar power is that their generation profile does not always align with demand. At times, RES may generate more power than the grid can absorb, forcing operators to curtail this clean energy. BESS can store this excess generation that would otherwise be lost and dispatch it later when it is needed, thereby increasing the total amount of renewable energy utilized and improving the economic return of RES assets \cite{b3}.
\end{enumerate}

\textbf{Grid Support and Resiliency Services:} This category includes services that enhance the overall robustness and security of the power system.

\begin{enumerate}
\item \textbf{Black Start Capability:} In the event of a major grid collapse or blackout, a BESS can provide the initial power required to restart major generation plants and systematically restore the entire grid. This "black start" capability is a critical function for ensuring system resilience\cite{b12}.
\item \textbf{Islanding and Microgrid Support:}
BESS enables microgrids to operate autonomously during large-scale outages, maintaining supply to critical loads such as hospitals and emergency facilities.
\item \textbf{Contingency and Spinning Reserve Support:}
Batteries can instantly deliver stored energy to replace lost generation capacity, reducing reliance on traditional spinning reserves.
\end{enumerate}

A critical examination of these services reveals a fundamental tradeoff between their operational requirements and the physical longevity of the battery. The operational profile demanded by fast-response services is starkly different from that of bulk-energy services. Frequency regulation, for instance, necessitates frequent, shallow, and high-power charge and discharge cycles to continuously track the grid's frequency signal \cite{b19}\cite{b20}. In contrast, a service like peak shaving typically involves one deep, slow, full charge-discharge cycle per day \cite{b21} \cite{b22}. Since battery degradation is a complex function of factors including cycle count, depth-of-discharge (DoD), C-rate (the rate of charge/discharge), and temperature, the choice of service portfolio directly dictates the battery's degradation pathway \cite{b23}. A control strategy optimized for frequency regulation will impose a high number of cycles, accelerating cycle-based aging mechanisms. A strategy focused on peak shaving will impose high stress from deep discharges, accelerating calendar and DoD-based aging. This establishes an undeniable causal chain: the selected service Portfolio dictates the Operational Profile, which in turn determines the Degradation Trajectory, and ultimately defines the asset's Lifecycle Cost through its impact on the battery replacement schedule\cite{b6}. The decision of which services to provide is therefore not merely an economic one; it is a profound technical decision with long-term consequences for the health and financial viability of the asset.

Furthermore, the economic value assigned to these ancillary services is directly linked to their technical specifications, which creates a powerful market-driven force shaping BESS technology itself. Markets for high-value, fast-response services, such as frequency regulation, exist because of the critical need for grid stability, and they often offer substantial revenue potential \cite{b3}. To qualify for participation in these markets, a BESS must meet stringent technical prerequisites, including millisecond-level response times and the ability to deliver high power outputs\cite{b14}. These demanding requirements inherently favor certain advanced battery chemistries, such as Lithium-ion, over older technologies like lead-acid, and necessitate sophisticated power conversion systems \cite{b14}. Consequently, the existence and regulatory structure of these ancillary service markets directly influence and justify investment in more advanced—and often more expensive—BESS hardware and control systems. The economic opportunity presented by the market makes the corresponding technical investment rational and necessary. Table \ref{tab:bess_technologies} compares the different BESS technologies and their respectively capability in terms of support to the power grid operation.

\begin{table*}[t]
\centering
\scriptsize
\caption{Comparison of Battery Energy Storage System (BESS) Technologies}
\begin{tabular}{|p{4cm}|p{1cm}|p{1cm}|p{1cm}|p{1cm}|p{1cm}|p{1cm}|p{1cm}|p{1cm}|}
\hline
\textbf{BESS Technology} & \textbf{Power Rating (MW)} & \textbf{Response Time (ms--s)} & \textbf{Lifetime (Cycles)} & \textbf{Efficiency (\%)} & \textbf{Cost (\$/kW)} & \textbf{Cost (\$/kWh)} & \textbf{Fast-Response Suitability} & \textbf{Bulk-Energy Suitability} \\
\hline
Lithium-Ion (Li-ion) & 0--5 & ms--s & 1000--3500 & 85--95 & 1200--4000 & 300--1300 & High & High \\
\hline
Lead-Acid            & 0--20 & ms--s & 500--1000  & 72.5--85 & 200--300   & 120--150  & Medium & Medium \\
\hline
Sodium-Sulfur (NaS)  & 0.05--8 & ms--s & 2000--4000 & 72.5--86 & 1000--3000 & 300--500  & High & High \\
\hline
Nickel-Cadmium (NiCd)& 0--40 & ms--s & 2000--3500 & --       & 500--1500  & 800--1500 & High & Medium \\
\hline
Vanadium Redox Flow (VRFB) & -- & ms--s & $>$10,000 & 75--85 & --         & --        & Medium & High \\
\hline
\end{tabular}
\label{tab:bess_technologies}
\end{table*}

\section{Section 2: Foundations of BESS Control and Management}

At the heart of every BESS lies a sophisticated control system responsible for translating high-level economic objectives and grid service requests into low-level physical actions, all while safeguarding the battery asset. This function is primarily carried out by the Battery Management System (BMS), an integrated hardware and software platform that acts as the operational brain of the battery pack \cite{b1}. A deep understanding of the BMS and its core functions is essential, as its performance and the accuracy of its estimations form the bedrock upon which all higher-level energy management and economic optimization strategies are built.

\subsection{The Central Role of the Battery Management System (BMS)}

The BMS is an electronic device that interfaces directly with the battery cells, the power conversion system, and the overarching Energy Management System (EMS). Its principal responsibilities are to enhance battery performance, ensure operation within safe limits, and ultimately extend the battery's useful life \cite{b23}. A comprehensive review in \cite{b24} outlines the key functionalities of a modern BMS, which can be categorized into four domains:
\begin{itemize}
\item Measurement and Monitoring: This is the most fundamental function. The BMS utilizes an array of sensors to continuously acquire real-time data on critical parameters for every cell or module in the pack, including voltage, current, and temperature. This raw data is digitized and serves as the primary input for all subsequent protection, control, and estimation functions.
\item Protection: The BMS acts as the primary defense mechanism against hazardous operating conditions. It actively monitors for and prevents events such as over-voltage (during charging), under-voltage (during discharging), over-current, and thermal runaway. If a dangerous threshold is approached, the BMS can take protective action, such as opening contactors to disconnect the battery pack from the load or charger, thereby preventing catastrophic failure and ensuring safety.
\item Computational Functions: This category encompasses the intelligent tasks performed by the BMS. State Estimation - This is arguably the most critical and complex computational function. The BMS must estimate key internal states of the battery that cannot be measured directly. The two most important states are:
\item State of Charge (SoC): SoC represents the current energy level of the battery as a percentage of its total capacity. Accurate SoC estimation is vital for any dispatch strategy, as it determines how much energy is available to be discharged or how much room is available for charging.
\item State of Health (SoH): SoH is a measure of the battery's condition relative to its brand-new state. It quantifies the extent of irreversible degradation the battery has experienced over its life. SoH is typically expressed in terms of two key parameters: capacity fade (the reduction in the total amount of charge the battery can store) and power fade as a function of internal resistance increase (quantifies reduction in the power capability and efficiency of the BESS). SoH is the single most important indicator of battery aging and is the primary determinant of the battery's remaining useful life (RUL) \cite{b25}\cite{b26}\cite{b27}.
\item Cell Balancing: In a large battery pack composed of hundreds or thousands of individual cells, minor manufacturing variations and temperature gradients can cause some cells to charge and discharge slightly faster than others. Over time, this leads to an imbalance in SoC across the pack. The BMS employs cell balancing circuits (either passive, which bleeds excess charge from higher-voltage cells, or active, which transfers charge from higher- to lower-voltage cells) to maintain uniformity. This is crucial for maximizing the usable capacity of the entire pack and preventing the premature aging of individual cells due to over-stress.
\item Communication: The BMS must communicate with external systems. It reports all measured and estimated data (SoC, SoH, temperature, etc.) to the higher-level EMS and receives dispatch commands in return. This communication, often via protocols like CAN bus or DNP3, is the link that connects the physical battery to the overall energy management strategy.
\end{itemize}

\begin{figure*}[!h]
	\centering
	\includegraphics[scale=0.55]{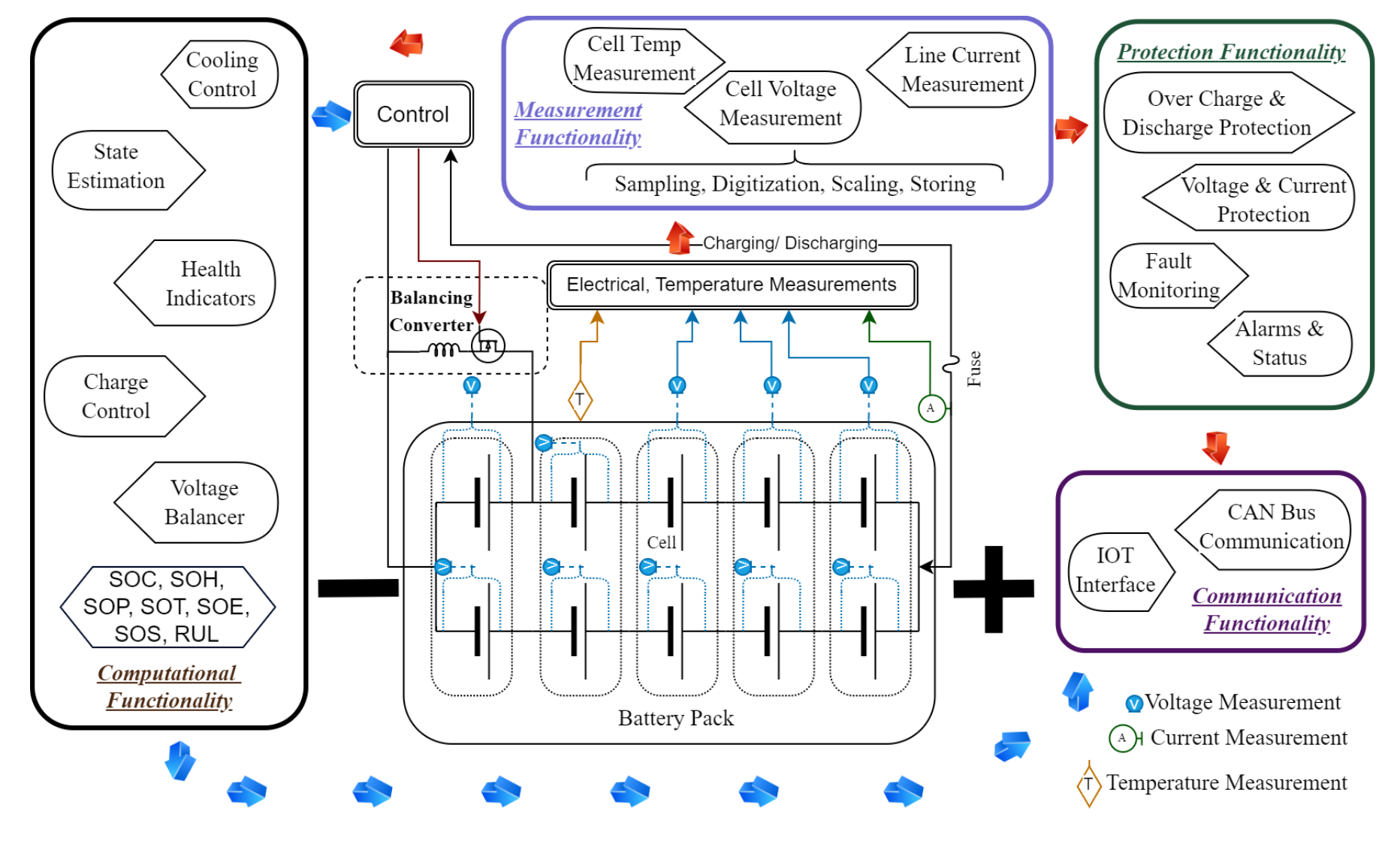}
	\caption{BESS system BMS architecture \cite{b9}}
	\label{fig:BESS BMS Arch}
\end{figure*}

Figure \ref{fig:BESS BMS Arch} presents the architecture of the BESS BMS system. The measurements for this system include temperature voltage and current values in real time from individual cells. The BMS algorithm is designed to maximize cell life while ensuring BESS life through a combination of BESS charge discharge commands.

\subsection{Health Indicators (HIs) as the Basis for SoH Estimation}

The estimation of SoH is not a direct measurement but an inference based on observable changes in the battery's behavior over time. The features used to make this inference are known as Health Indicators (HIs). The work summarized in \cite{b27} provides an exhaustive taxonomy of HIs, which can be broadly classified into two groups:
Primarily Measured HIs: These are features extracted directly from the time-series data of voltage, current, and temperature during charge-discharge cycles. Examples include the time it takes to charge at a constant current (CCCT), the voltage rise over a fixed time interval (VRET), or the highest temperature reached during discharge (HDT).
Calculated HIs: These are more complex features derived from the measured signals, often involving calculus. Examples include the slope of the voltage curve during charging (dV/dt) or the integral of the voltage curve (representing the area under the curve), which relates to the total energy transferred.
These HIs act as fingerprints of the battery's degradation state. As the battery ages, its internal chemistry changes, which manifests as subtle shifts in these indicators. By tracking these shifts, data-driven models (such as machine learning algorithms) within the BMS or a cloud analytics platform can accurately estimate the current SoH.

The quality of this estimation process has profound and often underappreciated implications for the entire BESS project. The accuracy of SoH estimation is a critical upstream dependency for any valid techno-economic analysis. The logic proceeds as follows: any credible techno-economic model must project revenues and costs over the project's entire lifecycle \cite{b28}. One of the most significant, if not dominant, costs is the eventual replacement of the battery pack \cite{b29}. The timing of this replacement is not arbitrary; it is triggered when the battery's SoH declines to a contractually defined warranty limit or an operational threshold (e.g., 80\% of original capacity). Simultaneously, the revenue-generating capability of the BESS is a direct function of its current SoH; a degraded battery has less capacity to sell and lower power capability \cite{b27}. Therefore, an inaccurate SoH estimation, as warned against in \cite{b27}, initiates a cascade of compounding errors throughout the financial model. An erroneously optimistic SoH reading will lead to an incorrect (delayed) prediction of the replacement date, which in turn leads to a flawed (underestimated) calculation of the lifecycle cost. This also results in an overestimation of future revenue-generating potential. The cumulative effect is a dangerously inflated Net Present Value (NPV) or an artificially low Levelized Cost of Storage (LCOS). This establishes a direct and powerful causal chain:

The Accuracy of SoH Estimation determines the Validity of the Degradation Model, which dictates the Accuracy of Lifecycle Cost and Revenue Projections, and ultimately underpins the Reliability of the Entire Techno-Economic Assessment. An error of just a few percentage points in SoH estimation can propagate into a massive error in the projected financial return over a 10- or 20-year project lifetime, turning a seemingly profitable project into a financial liability.

\section{Part II: The Economic Imperative: Market Participation and Financial Viability}

While the operational mandate focuses on what a BESS can and must do for the grid, the economic imperative addresses the equally important question of how these actions translate into value. A BESS is not just a technical solution; it is a significant capital investment that must provide a return. This part of the review shifts the focus from the physical asset to the economic ecosystem in which it operates, exploring the methodologies for assessing financial viability and the optimization strategies used to maximize economic benefit.

Section 3: Techno-Economic Analysis of BESS Deployments

Techno-economic analysis is the process of evaluating the financial viability of a BESS project by systematically accounting for all costs and revenues over its lifetime. It is the primary tool used by investors, developers, and policymakers to determine whether a BESS deployment is a sound financial decision.

\subsection{BESS Cost Structure}

A comprehensive techno-economic model must begin with a detailed breakdown of the project's cost structure. Drawing from frameworks presented in sources like \cite{b6}, the total lifecycle cost of a BESS can be deconstructed into three main categories:
Capital Expenditure (CAPEX): This represents the initial, one-time investment required to build the BESS project. It is the most visible cost component and includes:
The Battery System: This is often the largest single cost item, encompassing the battery cells, modules, and racks. The cost of lithium-ion batteries, while still significant, has fallen dramatically over the past decade, a trend that has been the single biggest driver of BESS adoption worldwide\cite{b23}.
Power Conversion System (PCS): This includes the inverters and converters that interface the DC battery with the AC grid.

Balance of Plant: This category includes the Battery Management System (BMS), thermal management systems (e.g., HVAC), switchgear, transformers, control systems, and the physical enclosure or container.
Soft Costs: These include costs for land acquisition, engineering, procurement, construction (EPC), permitting, and grid interconnection studies.

Operational Expenditure (OPEX): These are the recurring costs associated with running the BESS project over its lifetime. OPEX includes routine maintenance, software licensing fees, insurance, and the cost of charging the battery if it purchases energy from the grid for services like arbitrage or regulation.

Replacement Cost: This is a critical and often underestimated component of the total lifecycle cost. Batteries are consumable assets with a finite lifespan, determined by their degradation characteristics. The cost of replacing the battery stack one or more times during the project's overall lifetime (e.g., a 20-year project may require a battery replacement after 10 years) must be factored into the economic analysis\cite{b30}. The replacement profile varies significantly by battery chemistry; for example, a lead-acid battery may require replacement much sooner than a lithium-ion or sodium-ion battery, which impacts the long-term economics even if its initial CAPEX is lower.

\subsection{Key Economic Metrics and Methodologies}

Once costs and potential revenue streams are identified, analysts use several standard financial metrics to evaluate the project's viability:
Net Present Value (NPV): This is the most common metric for assessing project profitability. NPV calculates the difference between the present value of all future cash inflows (revenues) and the present value of all cash outflows (costs), discounted at a specific rate to account for the time value of money. A positive NPV indicates that the project is expected to be profitable, while a negative NPV suggests it is not \cite{b28}.

Levelized Cost of Storage (LCOS): LCOS represents the average cost per unit of energy discharged from the BESS over its entire lifetime ($/MWh$). It is calculated by dividing the total lifecycle cost (including CAPEX, OPEX, and replacement costs) by the total energy discharged over the battery's life. LCOS is an extremely useful metric for comparing the cost-effectiveness of different storage technologies or projects on an apples-to-apples basis \cite{b23}.

Internal Rate of Return (IRR): The IRR is the discount rate at which the NPV of a project becomes zero. It represents the project's expected annualized rate of return. A project is generally considered viable if its IRR is higher than the company's minimum acceptable rate of return or cost of capital \cite{b29}\cite{b30}.

Payback Period: This is the length of time required for the cumulative cash inflows from a project to equal the initial investment. While simple to calculate, it is a less sophisticated metric than NPV or IRR because it does not account for the time value of money or cash flows that occur after the payback period \cite{b30}.

\begin{figure*}[!h]
	\centering
	\includegraphics[scale=0.55]{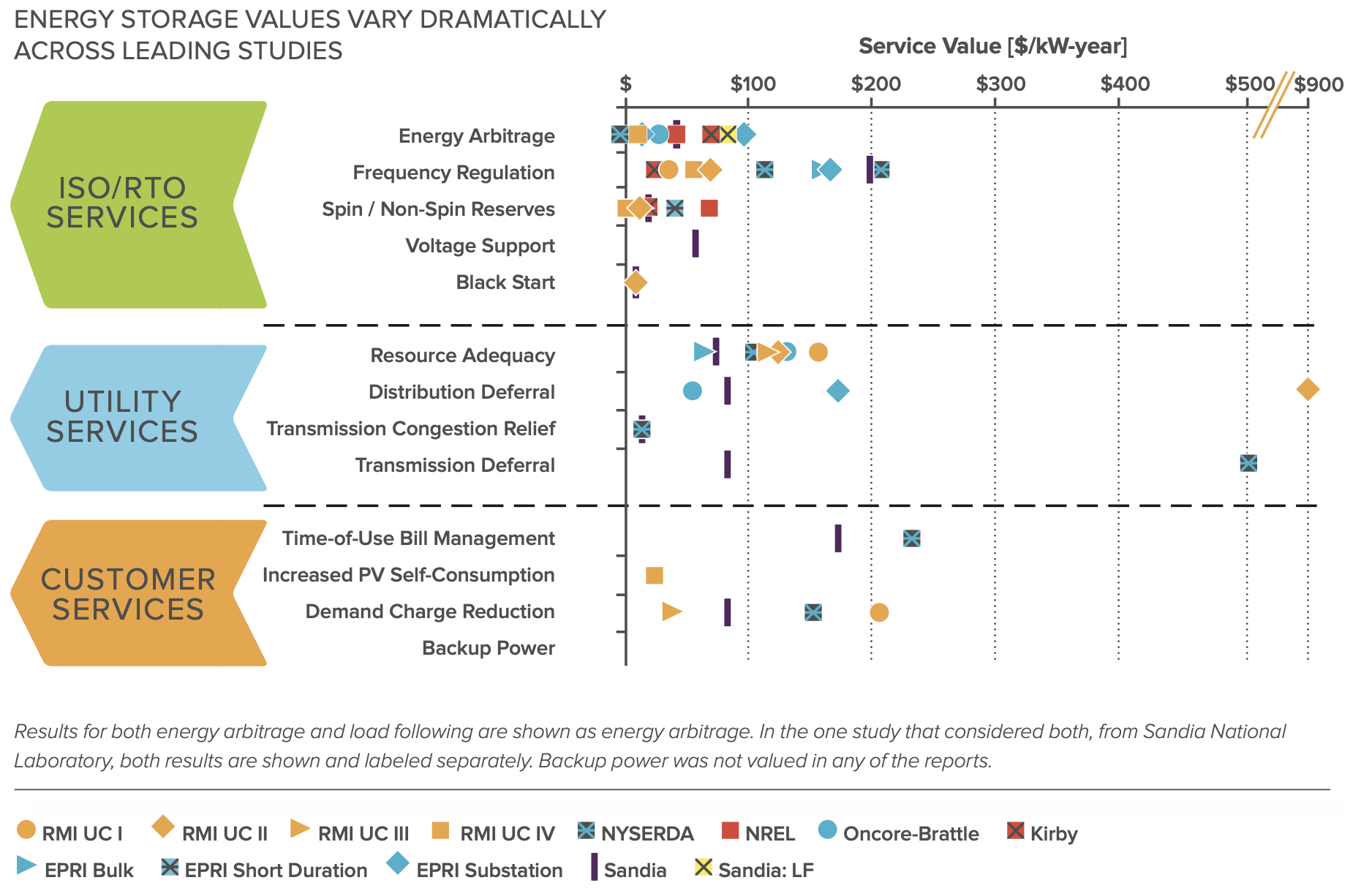}
	\caption{BESS services revenue \cite{b11}}
	\label{fig:BESS Revenue}
\end{figure*}

Figure \ref{fig:BESS Revenue} presents potential revenuw sources for a typical grid connected BESS system. These revenue sources form an important aspect in determining the economic feasibility and design of the BESS system.

\subsection{Case Study Findings and the Viability Gap}

The application of these techno-economic methodologies across the literature presents a decidedly mixed picture of BESS profitability. While there are specific scenarios where BESS projects demonstrate clear financial viability, a significant portion of the research concludes that standalone BESS deployments often struggle to be profitable without some form of external support.

For example, studies have shown that BESS can achieve a positive NPV when deployed in markets with very high demand charges (where reducing peak consumption yields large savings) or when they can participate in particularly lucrative ancillary service markets \cite{b28}. A techno-economic analysis for Swedish industries found that using a BESS for peak load shaving was only profitable for very small reductions (1-3\%); higher levels of peak shaving resulted in a negative NPV due to the rapidly increasing size and cost of the required BESS \cite{b28}. Another study found that behind-the-meter storage systems in California and Tennessee only yielded positive NPVs under specific utility rate structures and optimistic battery cost assumptions; most other scenarios were uneconomical \cite{b31}. A feasibility study for integrating a large-scale BESS with a synchronous generator concluded that the project was technically feasible and provided grid benefits, but the costs outweighed the anticipated revenues, leading to a negative NPV over the project's life\cite{b25}.

This body of evidence points to a significant contradiction that lies at the heart of the BESS challenge. On one hand, Part I of this review established the immense and growing technical value of BESS for enabling the energy transition and ensuring grid stability \cite{b12}.

The services they provide are becoming essential. On the other hand, these multiple techno-economic analyses demonstrate that the
economic value captured through existing market mechanisms is often insufficient to cover the high capital and replacement costs of the asset \cite{b25}.

This discrepancy creates a - viability gap: the power system needs the services that BESS provide, but the market structure does not always compensate those services at a level that makes the investment profitable on a standalone basis. This gap is the fundamental economic problem that the concept of synergy is intended to solve. Advanced strategies like revenue stacking and degradation-aware control are less mature from a commercial perspective; they are practical, commercial necessities developed to close this viability gap. They represent a concerted effort to extract the maximum possible value from every aspect of the BESS operation to make projects bankable and accelerate their deployment. The economic challenge, therefore, is the primary driver compelling the industry to move beyond siloed thinking and embrace a more intelligent, synergistic approach to BESS management.

\subsection{Section 4: Optimization Algorithms for BESS Economic Dispatch}

To bridge the viability gap and maximize the economic return from a BESS asset, operators rely on sophisticated optimization algorithms. These algorithms form the core of the BESS Energy Management System (EMS) and are responsible for making the critical moment-to-moment decisions about when to charge, discharge, or remain idle. The evolution of these algorithms, as documented extensively in the literature, provides a clear narrative of the field's progression from simple, deterministic problems to highly complex, dynamic, and uncertain real-world environments \cite{b19}.

\subsubsection{Evolution of Optimization Techniques}
The development of BESS dispatch algorithms reflects a direct response to the increasing complexity of both the market opportunities and the operational realities of the battery itself. This evolution can be traced through several distinct stages:
\begin{itemize}
\item Classical and Deterministic Optimization: The earliest and most foundational approaches to BESS scheduling employed classical optimization techniques. These include methods like Linear Programming (LP), Mixed-Integer Linear Programming (MILP), and Dynamic Programming (DP) \cite{b32}. These methods are highly effective for solving well-defined problems where all parameters, such as future electricity prices and load profiles, are known with certainty. For example, a simple energy arbitrage problem with a known 24-hour price forecast can be effectively solved using MILP to determine the optimal charge/discharge schedule that maximizes profit. While powerful and capable of finding a mathematically provable optimal solution, their primary limitation is their inability to handle uncertainty, a defining characteristic of modern energy markets.
\item Heuristic and Metaheuristic Algorithms: To address more complex, non-linear, and non-convex optimization problems for which classical methods are ill-suited, researchers turned to heuristic and metaheuristic algorithms. Prominent examples in the BESS literature include Particle Swarm Optimization (PSO) and Genetic Algorithms (GA).13 Unlike classical methods that require a precise mathematical formulation of the problem, metaheuristics use nature-inspired processes to intelligently search the solution space. They are adept at finding high-quality, near-optimal solutions for complex problems, such as the optimal sizing and placement of BESS in a distribution network, without getting trapped in local optima. Their trade-off is that they do not guarantee finding the absolute global optimum.
Stochastic and Robust Optimization: As the need to explicitly manage the uncertainty of real-time electricity prices, intermittent RES generation, and unpredictable load became paramount, the field advanced to stochastic and robust optimization frameworks \cite{b19}.
\end{itemize}

Stochastic optimization models uncertainty using probability distributions, aiming to find a dispatch strategy that maximizes the expected profit over a range of possible future scenarios. Robust optimization, in contrast, takes a more conservative approach, seeking a strategy that performs reasonably well under the worst-case realization of the uncertain parameters. These methods represent a significant step forward in creating dispatch strategies that are resilient to real-world volatility.

The Rise of Artificial Intelligence and Reinforcement Learning (RL): The most recent and arguably most transformative trend in BESS optimization is the application of artificial intelligence, particularly Reinforcement Learning (RL)\cite{b33}\cite{b34} RL represents a paradigm shift from the model-based approaches described above. In an RL framework, an agent (the BESS controller) learns the optimal dispatch strategy not from a pre-defined model, but through direct trial-and-error interaction with its environment (the real or, more commonly, a simulated electricity market). The agent takes actions (charge/discharge), observes the resulting state (SoC, market price) and receives a reward (profit or cost). Over many iterations, the agent learns a policy—a mapping from states to actions—that maximizes its cumulative long-term reward \cite{b35}\cite{b36}. This model-free approach makes RL exceptionally well-suited for the complex, dynamic, and highly uncertain environment of modern power systems. It can implicitly learn to balance multiple competing objectives (e.g., revenue maximization and degradation minimization) without needing an explicit mathematical model of their relationship. Studies have shown that RL-based energy management systems can improve the operational efficiency of BESS by more than 30\% compared to static, rule-based controllers \cite{b34}.

The progression from simple deterministic models to advanced AI agents is not merely a quest for better algorithms in isolation. It is a clear example of co-evolution, where the tools are evolving in direct response to the increasing complexity of the problem they are trying to solve. In a simple, predictable world with a single revenue stream and a perfect battery, a MILP model would suffice. However, as electricity markets evolve to include multiple, co-existing, and often conflicting revenue streams (ancillary services, capacity payments, real-time energy), the problem becomes a multi-objective one \cite{b19}. When the profound uncertainty of real-time prices and RES generation is layered on top, deterministic models become inadequate, necessitating the move to stochastic or adaptive approaches \cite{b26}. Finally, when the complex, non-linear, and path-dependent reality of battery degradation is incorporated as a key factor to be managed, the problem's dimensionality and complexity explode \cite{b4}.

It is at this final stage of complexity where RL demonstrates its true power. An RL agent can learn a sophisticated control policy that implicitly navigates the trade-offs between short-term profit, long-term degradation, and market uncertainty \cite{b36}\cite{b37}\cite{b38}\cite{b39}. Therefore, the evolution of these optimization algorithms is a direct reflection of the evolution of our understanding of the BESS management problem itself. RL is not just another alternative in the toolbox; it is increasingly becoming the necessary and most appropriate tool to solve the problem as it is now holistically defined. Table \ref{tab:algorithm_comparison_with_refs} presents a comparison of different optimization algorithms deployed for BESS energy management.

\begin{table*}[t]
\centering
\scriptsize
\caption{Taxonomy and Evolution of Optimization Algorithms for BESS Economic Dispatch}
\begin{tabular}{|p{3cm}|p{3cm}|p{3cm}|p{3cm}|p{2cm}|p{1cm}|}
\hline
\textbf{Algorithm Class} & \textbf{Primary Objective} & \textbf{Handling of Uncertainty (Prices, RES)} & \textbf{Inclusion of Battery Degradation Model} & \textbf{Computational Complexity} & \textbf{Key References} \\
\hline
Classical/Deterministic \newline (MILP, DP) & Cost/Profit Optimization & None (Deterministic forecasts) & Not included or highly simplified (e.g., fixed cycle life) & Low to Medium & [1], [2] \\
\hline
Heuristic/Metaheuristic \newline (PSO, GA) & Multi-objective Optimization \newline (e.g., sizing, placement) & Scenario-based (can handle variations) & Can be included as an objective or constraint & Medium & [5], [19] \\
\hline
Stochastic/Robust & Optimization under Uncertainty & Explicitly modeled (Probabilistic/Worst-case) & Can be included in the stochastic formulation & High & [16], [17] \\
\hline
AI/Reinforcement Learning \newline (Q-Learning, PPO) & Maximize Long-Term Reward & Model-Free / Adaptive Learning & Implicitly learned via reward function or explicitly as a cost & High (Training) \newline Low (Inference) & [35], [37] \\
\hline
\end{tabular}
\label{tab:algorithm_comparison_with_refs}
\end{table*}

Part III: Synthesis and Synergy: The Future of BESS Energy Management

Having established the distinct operational mandates and economic imperatives governing BESS, this final part of the review integrates these two perspectives. It moves beyond the siloed analyses of control and economics to explore the synergistic frameworks that represent the frontier of BESS energy management. This synthesis is the core of the argument for a new, holistic paradigm. It focuses on how the conflict between revenue and reliability can be transformed into a managed, optimized trade-off, and it identifies the emerging technologies that make this synergy possible in practice.

Section 5: The Nexus of Control and Economics: Degradation-Aware Revenue Stacking

The most direct path to closing the viability gap identified in Part II is to move beyond single-application operation and instead leverage the BESS's versatility to capture multiple streams of value. This strategy, known as revenue stacking or value stacking, is widely recognized in the literature as critical for improving BESS profitability \cite{b22}. However, implementing revenue stacking effectively requires a sophisticated energy management strategy that can navigate the inherent conflicts between different services and, most importantly, integrate a realistic understanding of battery degradation as a direct economic cost.

\subsection{Frameworks for Revenue Stacking}

A BESS can, in theory, participate in multiple markets, such as performing energy arbitrage in the day-ahead market while also holding capacity in reserve to provide frequency regulation if called upon. This approach can dramatically improve financial returns; one study found that revenue stacking could boost annual revenues by as much as 129\%, reducing the project payback period by an average of 8 years \cite{b40}\cite{b41}.

However, this strategy introduces significant operational complexity and potential conflicts. For example, if a BESS operator commits the full capacity of the battery to an energy arbitrage cycle (charging fully overnight to sell during the evening peak), that capacity is unavailable if a lucrative, unexpected opportunity to provide fast frequency response arises mid-day. Failing to provide a contracted service can result in significant financial penalties, negating any potential gains.

To manage these conflicts, researchers have proposed structured scheduling frameworks. The work detailed in \cite{b41} provides a particularly insightful example for a BESS in the Northern Ireland market. This framework addresses conflicts by prioritizing services and allocating specific time windows for each activity based on grid needs and market rules.

\begin{itemize}
\item Winter/Autumn Operation: The primary grid need is peak shaving in the evening. The framework prioritizes this by reserving the BESS for Distribution Network Support (DNS) from 17:00 to 22:00. To prepare, the BESS charges from the grid during the low-price overnight period (01:00-08:00). Any remaining capacity after the DNS period can be sold back to the market. This structure ensures the primary grid service is met, while secondary arbitrage opportunities are fit in around it.
\item Summer/Spring Operation: The primary grid need shifts to absorbing excess solar PV generation mid-day. The framework reserves the BESS for DNS charging from 10:00 to 15:00. To ensure the battery is empty and ready to absorb this solar energy, it is scheduled to perform a full arbitrage cycle in the early morning (charge 01:00-05:00, sell 05:00-10:00).
\end{itemize}

These frameworks demonstrate how to resolve service conflicts through intelligent, priority-based temporal scheduling. More advanced models take this a step further by co-optimizing dispatch across multiple services simultaneously, often on different timescales. For instance, the model in \ref{b32} co-optimizes energy arbitrage (EA) on a five-minute basis with frequency regulation (FR) on a two-second basis, creating a truly dynamic, multi-service dispatch strategy.

Modeling Battery Degradation as an Economic Cost:

The most crucial element of a truly synergistic framework is the integration of battery degradation not just as a physical constraint, but as a quantifiable economic cost within the optimization problem itself. This is the essential link that connects the control-centric view with the economic-centric view.

Without this integration, the relationship between revenue and degradation is a simple, unmanaged conflict: more usage leads to more revenue but also more degradation \cite{b29}. A naive control strategy might try to manage this by imposing hard operational limits (e.g., never discharge below 20\% SoC or above a C-rate of 1). While this protects the battery, it is a blunt instrument that is economically suboptimal, as it prevents the BESS from capturing high-value opportunities that may require temporarily exceeding these conservative limits.

\begin{figure*}[!h]
	\centering
	\includegraphics[scale=0.5]{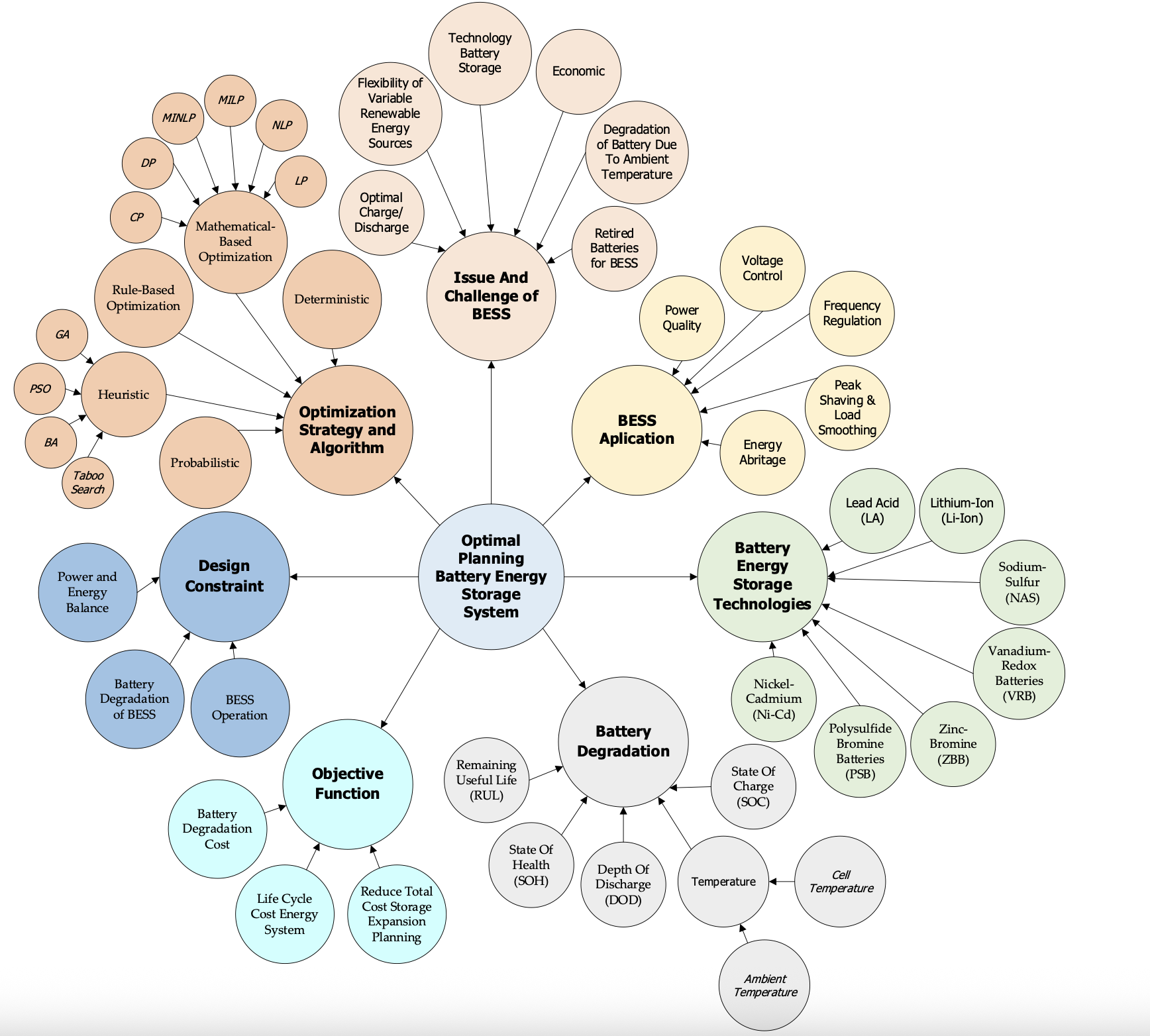}
	\caption{Minmap of BESS optimization [15]}
	\label{fig:BESS_Optimization}
\end{figure*}

A degradation-aware optimization algorithm reframes the entire problem. The methodology involves creating a cost function for degradation, where the marginal cost of a given charge or discharge action is calculated based on the amount of battery life it consumes \cite{b4}. This cost can be derived from sophisticated models that account for DoD, C-rate, temperature, and SoC. By assigning a dollar value to each increment of degradation, the operational goal of preserving life is transformed into an economic input of minimizing cost. Figure \ref{fig:BESS_Optimization} provides a description of different BESS dispatch strategies being used in production systems.

The impact of this reframing is profound. The optimization algorithm is no longer making a binary decision based on hard limits; it is making a continuous economic trade-off. For every potential dispatch action, it asks the question: Is the marginal revenue I will gain from this action greater than the marginal cost of the degradation it will cause? This allows the system to make economically rational decisions on a moment-by-moment basis. For example:

\begin{itemize}
\item It might decide to perform a high-stress, deep discharge cycle if the electricity price is extremely high, because the high revenue justifies the high degradation cost.
\item It would, however, forego the exact same physical action if the price signal were only moderate, because the revenue would not be sufficient to cover the degradation cost.
\end{itemize}

This approach leads to more conservative but ultimately more profitable long-term operational strategies. As demonstrated in studies like \cite{b32} and \cite{b4}, including degradation in the optimization model can significantly reduce the rate of battery aging \cite{b32} without a correspondingly significant loss in short-term profit. The small revenues that are sacrificed by avoiding high-stress cycles are more than compensated for by the extended life of the asset and the deferral of a very costly battery replacement.

This synergistic optimization, therefore, fundamentally transforms the relationship between revenue generation and battery health. It moves the system away from a zero sum conflict, where one comes at the expense of the other, and toward a managed, quantifiable trade off. The synergy is not in eliminating degradation, which is an unavoidable consequence of use, but in ensuring that every unit of the battery's finite life is spent in the most economically advantageous way possible. This maximizes the total lifetime value of the asset, aligning the goals of the engineer and the economist into a single, unified objective function.

\noindent
The synergistic optimization framework can be mathematically expressed as an objective function that maximizes the total lifetime value of the battery asset:

\begin{equation}
J = \int_{t_0}^{t_\mathrm{end}} \Big[ R(t) - \lambda \, C_\mathrm{deg}(t) \Big] \, dt
\label{eq:objective}
\end{equation}

subject to operational, market, and hardware constraints.

{Definition of terms:}
\begin{itemize}
    \item $J$ -- Total lifetime economic value of the battery asset. This represents the cumulative net benefit derived from all operations over the asset's usable life.
    \item $R(t)$ -- Instantaneous revenue rate [\$/h] at time $t$, incorporating income from energy arbitrage, ancillary services, capacity markets, and other compensated grid functions.
    \item $C_\mathrm{deg}(t)$ -- Instantaneous degradation cost rate [\$/h], representing the monetized wear of the battery due to cycling, depth of discharge, and thermal stress. It can be modeled as:
    \begin{equation}
        C_\mathrm{deg}(t) = k_\mathrm{deg} \, \dot{D}(t)
    \end{equation}
    where $\dot{D}(t)$ is the rate of degradation (e.g., capacity loss rate) and $k_\mathrm{deg}$ is the cost per unit of degradation [\$/\% capacity loss].
    \item $\lambda$ -- Trade-off weighting factor that quantifies the value placed on battery longevity relative to immediate revenue. A higher $\lambda$ emphasizes life preservation; a lower $\lambda$ prioritizes short-term profitability.
\end{itemize}
The cost function in Eq.~\ref{eq:objective} formalizes the balance between short-term revenue and long-term asset health. Rather than treating degradation as an unavoidable loss, the formulation explicitly prices it within the control objective. The controller thus dispatches the BESS only when the marginal revenue $R(t)$ exceeds the degradation cost $\lambda C_\mathrm{deg}(t)$, ensuring that each unit of the battery’s finite lifetime is utilized in the most economically advantageous manner. This transforms the conventional zero-sum conflict between profitability and durability into a managed, quantifiable trade-off, uniting the objectives of the engineer and the economist within a single, optimized framework.

\section{Section 6: Digital Twins as the Ultimate Synergistic Framework}

The theoretical frameworks for degradation-aware, multi-service optimization are powerful, but their practical implementation requires a platform that can handle immense complexity, integrate diverse data streams, and operate in real time. This is the role of the Digital Twin (DT), a technology that represents the culmination of the synergistic approach. A DT is not merely a simulation or a monitoring dashboard; it is a living, high-fidelity virtual replica of a physical BESS, continuously synchronized with its real-world counterpart through a constant flow of data \cite{b42}.

Architecture of BESS Digital Twins

A comprehensive BESS Digital Twin, such as the framework proposed in \cite{b43} and \cite{b44}, is a multi-layered, multi-physics system. Its architecture includes several key components:
\begin{itemize}
\item Multi-Scale Virtual Representation: The core of the DT is a set of interconnected models that replicate the BESS at every level of granularity. This starts with intricate multiphysics models at the individual cell level, capturing the electrochemical and thermal behavior. These cell models are then aggregated to create models of modules, packs, and racks, and finally integrated with models of the auxiliary components like the PCS and the thermal management system. This hierarchical structure ensures a precise and holistic virtual representation of the entire BESS\cite{b44}.
\item Interface and Data Layer: A robust interface layer is responsible for the bidirectional flow of information between the physical asset and its digital counterpart. This includes a data acquisition system for collecting sensor data (voltage, current, temperature) in real time, data quality assurance protocols to ensure the data is clean and reliable, and strong cybersecurity measures to protect the system from unauthorized access and manipulation \cite{b44}.
\item Analytics Engine: The DT incorporates an advanced analytics engine that leverages the real-time data and the virtual models to generate actionable intelligence. This engine performs three types of analytics:
\item Descriptive Analytics: Interpreting historical and current data to understand what is happening with the BESS right now.
\item Predictive Analytics: Forecasting future states, such as the trajectory of SoH degradation or the Remaining Useful Life (RUL) of the battery.
\item Prescriptive Analytics: Recommending optimal actions, such as adjusting the dispatch strategy or scheduling maintenance, to improve performance and mitigate risks.
\end{itemize}

\begin{figure}[!h]
	\centering
	\includegraphics[scale=0.5]{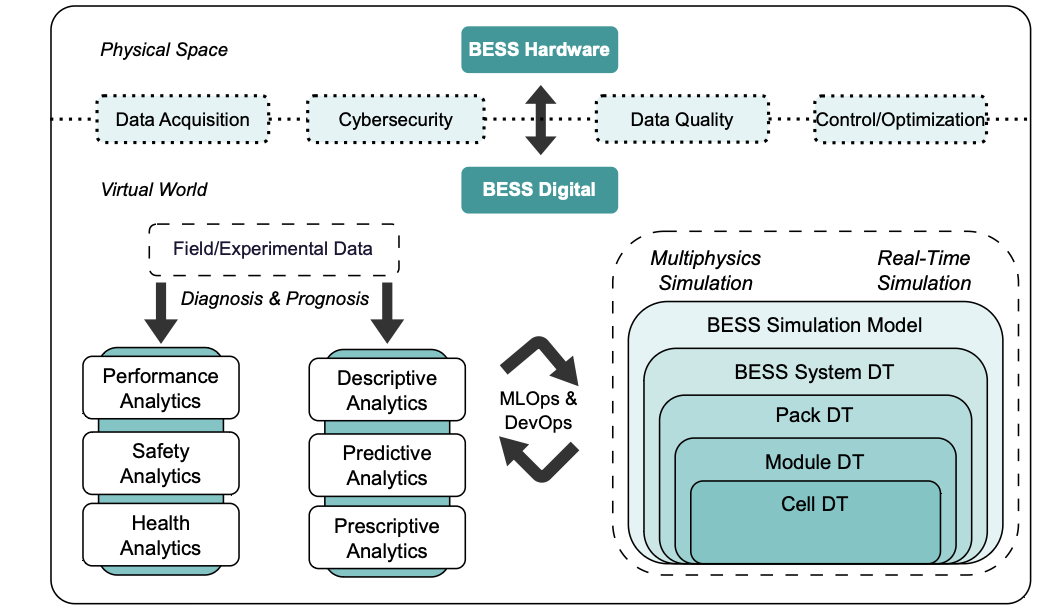}
	\caption{Digital Twin for BESS [44]}
	\label{fig:Digital Twin BESS}
\end{figure}

Figure \ref{fig:Digital Twin BESS} presents the architecture of BESS digital twin from \cite{b44}. The data from physical BESS is used as input to update the state of the digital twin. A optimization algorithm in this system used to both analyze and predict future state followed by a decision to either charge or discharge the BESS system.

\subsection{DT-Enabled Applications for Synergy}

The true power of the Digital Twin lies in its ability to serve as the central nervous system for the synergistic management of the BESS. It enables applications that were previously impractical or impossible.
High-Fidelity Predictive Maintenance and SoH Forecasting: Traditional SoH estimation methods often rely on simplified models or infrequent laboratory-style tests. A DT, in contrast, combines the live operational data from the physical BESS with its sophisticated, physics-based degradation models. This allows it to generate a continuously updated, highly accurate estimate of the battery's true SoH and a reliable forecast of its RUL \cite{b27}. This capability directly addresses the critical need for accurate SoH estimation that was identified as the bedrock of any valid techno-economic analysis (Insight 3), replacing assumptions with data-driven reality.

\begin{figure*}[!h]
	\centering
	\includegraphics[scale=0.5]{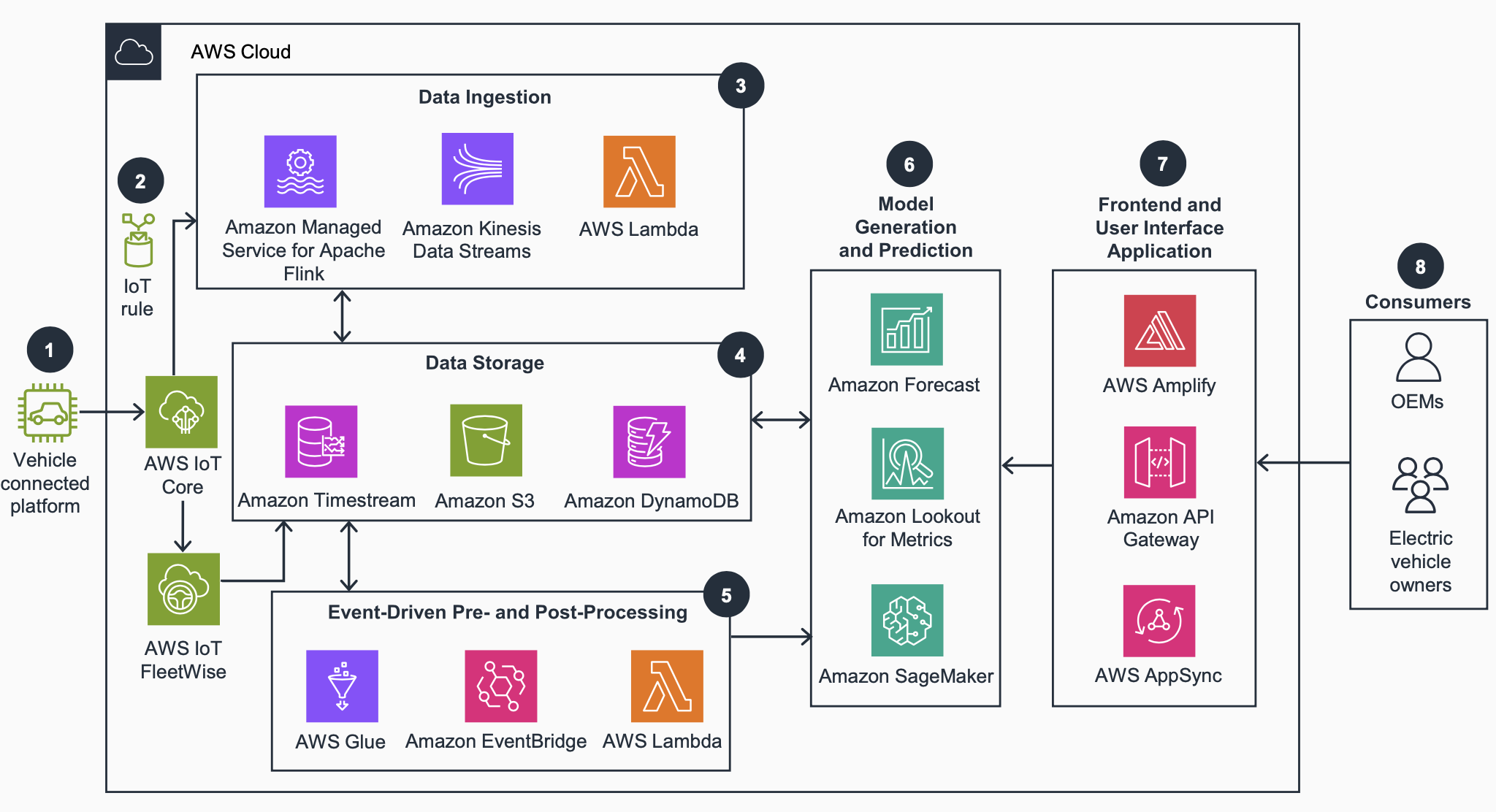}
	\caption{Digital Twin in AWS Cloud [43]}
	\label{fig:AWS_BESS_DT}
\end{figure*}

Holistic Techno-Economic Optimization: The DT provides the ideal sandbox and real-time data source for the advanced, degradation-aware optimization algorithms discussed in the previous section. It allows operators to:

Simulate and De-Risk Strategies: Before deploying a new revenue-stacking strategy on the physical asset, operators can run simulations on the DT to accurately predict its impact on both revenue and long-term battery health. This allows for the comparison of multiple strategies to find the optimal balance without risking the physical asset \cite{b45}\cite{b46}

Incorporate Real-Time Health into Dispatch: The highly accurate, real-time SoH and internal resistance estimates generated by the DT can be fed directly into the economic dispatch algorithm as key parameters. This means the degradation cost used in the optimization is not a static assumption but a live value reflecting the battery's actual, current condition \cite{b44}

Close the Optimization Loop: The DT enables a closed-loop optimization process where the system continuously balances aggressive market participation against long-term asset integrity. The DT can identify that a certain dispatch strategy is causing faster-than-expected degradation and prescriptively recommend a modification to the strategy to bring the degradation back in line with the long-term financial model \cite{b47}\cite{b48}.

A Digital Twin should not be viewed as merely an advanced monitoring tool. It is the functional embodiment of the control-economic synergy. It achieves this by creating a powerful and continuous feedback loop that was previously missing. The process works as follows:
\begin{enumerate}  
\item The Physical BESS operates in the real world, and its sensors stream real-time data (voltage, current, temperature) to its virtual counterpart \cite{b44}
\item This data is ingested by the Digital Twin. Its physics-based models process this data to update their internal state, generating a highly accurate, real-time estimate of the battery's true SoH, internal resistance, capacity fade, and other critical degradation parameters.27
\item This high-fidelity health status is then fed from the DT back to the Economic optimization engine (the EMS). This live data is used to calculate the real-time marginal cost of degradation for any potential dispatch action \cite{b44}\cite{b47}\cite{b48}.
\item The EMS solves the optimization problem, balancing market revenue against this real-time degradation cost, and issues a new, holistically optimized dispatch command back to the Physical BESS.
\end{enumerate}

Figure \ref{fig:AWS_BESS_DT} presents a digital twin architecture being used for production EV's to manage their charge and discharge rate. A similar architecture can be used to enable the proposed digital twin from this work in a production ready environment. Architectures such as the one from this Figure provide a highly scalable compute solution that can be enabled across potentially hundreds of BESS systems deployed in a given power grid while ensuring low cost and high reliability of the BESS EMS systems.

This continuous Physical-to-Digital-to-Economic-to-Physical feedback loop is a critical pathway to achieve the synergy required for the BESS systems. The Digital Twin acts as the critical bridge, the translator that allows the complex, non-linear, and stochastic physical reality of the battery to be accurately represented, priced, and managed within the rational, economic decision-making framework. It moves BESS management from a world of static optimization based on assumptions and historical data to a new paradigm of dynamic, adaptive optimization based on real-time, physical truth.

\section{Conclusion and Future Research Directions}
\subsection{Recapitulation of Key Synergies}
This analytical review has charted the evolution of Battery Energy Storage System management from a field characterized by a dichotomy between operational control and economic analysis to an emerging paradigm defined by their deep and necessary synergy. The evidence drawn from the literature demonstrates that the immense technical value of BESS in modernizing the power grid can only be translated into economic viability through integrated management frameworks. The central theme of this review has been built upon a series of interconnected conclusions.
\begin{enumerate}

\item First, a fundamental trade-off exists between the operational demands of delivering various ancillary services and the resulting physical degradation of the battery asset. Fast-response services like frequency regulation impose high cycle counts, while bulk-energy services like peak shaving impose high depth-of-discharge stress, each accelerating different aging pathways. This establishes a direct causal link between the selected service portfolio and the asset's lifecycle cost.

\item Second, to justify the high capital cost of BESS, operators are driven toward multi-service participation and revenue stacking. However, this creates complex scheduling challenges and potential conflicts that require intelligent, priority-based management frameworks to resolve.

\item Third, the most critical step toward synergy is the re framing of battery degradation from a purely physical constraint into a quantifiable economic cost within the optimization algorithm. This transforms the relationship between revenue and longevity from a zero-sum conflict into a manageable, dynamic trade-off, allowing the system to ensure that every unit of battery life is expended for the maximum possible economic return.

\item Finally, Digital Twin technology emerges as a key enabling platform for this synergistic vision. By creating a high-fidelity, living virtual replica of the physical asset, the DT provides the foundation for superior predictive maintenance and, most importantly, creates a continuous Physical-to-Digital-to-Economic-to-Physical feedback loop. This loop allows the complex, real-time physical state of the battery to dynamically inform its economic dispatch strategy, moving the field from static optimization based on assumption to dynamic optimization based on reality.
  
\end{enumerate}

\subsection{Identification of Research Gaps}

While the path toward fully synergistic BESS management is clear, the review of the current literature also highlights several significant research gaps that must be addressed to accelerate progress:

\begin{enumerate}
\item Standardization of Digital Twin Models and Data Frameworks: While many researchers are developing DTs for BESS, there is a distinct lack of standardized frameworks, data models (like DTDL), and communication protocols\cite{b49}. This fragmentation hinders interoperability between systems from different vendors, makes it difficult to compare the performance of different DT solutions, and slows the development of a robust ecosystem of third-party analytics and optimization services.
\item Market Design and Regulatory Barriers: Current ancillary service markets were often designed for the capabilities of traditional thermal generators and may not be structured to properly value the unique, fast-acting capabilities of BESS. Regulations can create barriers to multi-service participation, and market products may not be defined in a way that allows for the efficient stacking of revenues \cite{b38} \cite{b39}. Further research is needed into market redesigns that explicitly enable and incentivize the synergistic, multi-service operation of energy storage.
\item Computational Complexity and Real-Time Execution: The vision of a high-fidelity, multi-physics Digital Twin running in lockstep with a complex, AI-based optimization algorithm presents a formidable computational challenge \cite{b9}. Ensuring that these computationally intensive models can execute in real time to make dispatch decisions on a timescale of seconds or less remains a significant hurdle. Research into model order reduction, computationally efficient AI architectures, and distributed or edge computing solutions is needed to make this vision practical for widespread deployment \cite{b50}\cite{b51}. 
\end{enumerate}

\subsection{Future Outlook}

The future of BESS management lies in the continued development and integration of these synergistic principles, culminating in the creation of fully autonomous energy management systems. These systems will leverage AI-powered Digital Twins as their cognitive core. They will autonomously scan market opportunities across a spectrum of energy, capacity, and ancillary service markets; use their internal, high-fidelity models to predict the revenue potential and associated degradation cost of each opportunity; and self-optimize their dispatch strategy in real time to maximize lifetime value. By continuously balancing the competing objectives of revenue generation, grid support, and asset longevity, these intelligent systems will be capable of delivering the lowest possible Levelized Cost of Storage while simultaneously enhancing the reliability and resilience of the power grid. This represents the ultimate goal of BESS Energy Management Systems: to transform the battery from a passively controlled device into an autonomous, value-seeking economic agent that is a cornerstone of a clean, efficient, and secure energy future.

\section*{Acknowledgment}
The authors would like to thank the Advanced Algorithms team at SouthWest Research Institute who contributed to various aspects of this research.

\end{document}